\documentclass[twocolumn,prb,showpacs,preprintnumbers,superscriptaddress]{revtex4}
\usepackage{mathptm}
\usepackage{dcolumn}
\usepackage{enumerate}
\usepackage{graphicx}

\begin{document}
\title{Why MgFeGe is not a superconductor}
\author{Harald O. Jeschke}
\affiliation{Institut f\"ur Theoretische Physik, Goethe-Universit\"at Frankfurt am Main, 60438 Frankfurt am Main, Germany}
\author{I. I. Mazin}
\affiliation{Naval Research Laboratory, 4555 Overlook Ave. SW, Washington, DC 20375}
\author{Roser Valent{\'\i}}
\affiliation{Institut f\"ur Theoretische Physik, Goethe-Universit\"at Frankfurt am Main, 60438 Frankfurt am Main, Germany}
\date{\today}

\begin{abstract}
  The recently synthesized MgFeGe compound is isostructural and
  isoelectronic with superconducting LiFeAs. Both materials are
  paramagnetic metals at room temperature. Inspection of their
  electronic structures without spin polarization reveals hardly any
  difference between the two. This fact was interpreted as evidence
  against popular theories relating superconductivity in Fe-based
  materials with spin fluctuations. We show that in the magnetic
  domain the two compounds are dramatically different, and the fact
  that MgFeGe does not superconduct, is, on the contrary, a strong
  argument in favor of theories based on spin fluctuations.
\end{abstract}

\maketitle

Soon after the discovery of iron based
superconductors~\cite{Kamihara2008}, spin fluctuations have been
proposed as the pairing glue~\cite{Mazin2008}. So far, many
experiments are compatible with or even supportive of a spin
fluctuation mediated mechanism~\cite{Stewart2011,Hirschfeld2011}. In
the weak coupling approach to this pairing scenario, the nesting
properties of the Fermi surface are important. Therefore, two
materials with virtually identical Fermi surfaces and electron count,
where one of the materials superconducts and the other doesn't, would
at first glance cast a doubt on the validity of such an approach to
superconducting pairing.

\begin{figure}[ptb]
\includegraphics[width=0.95\columnwidth]{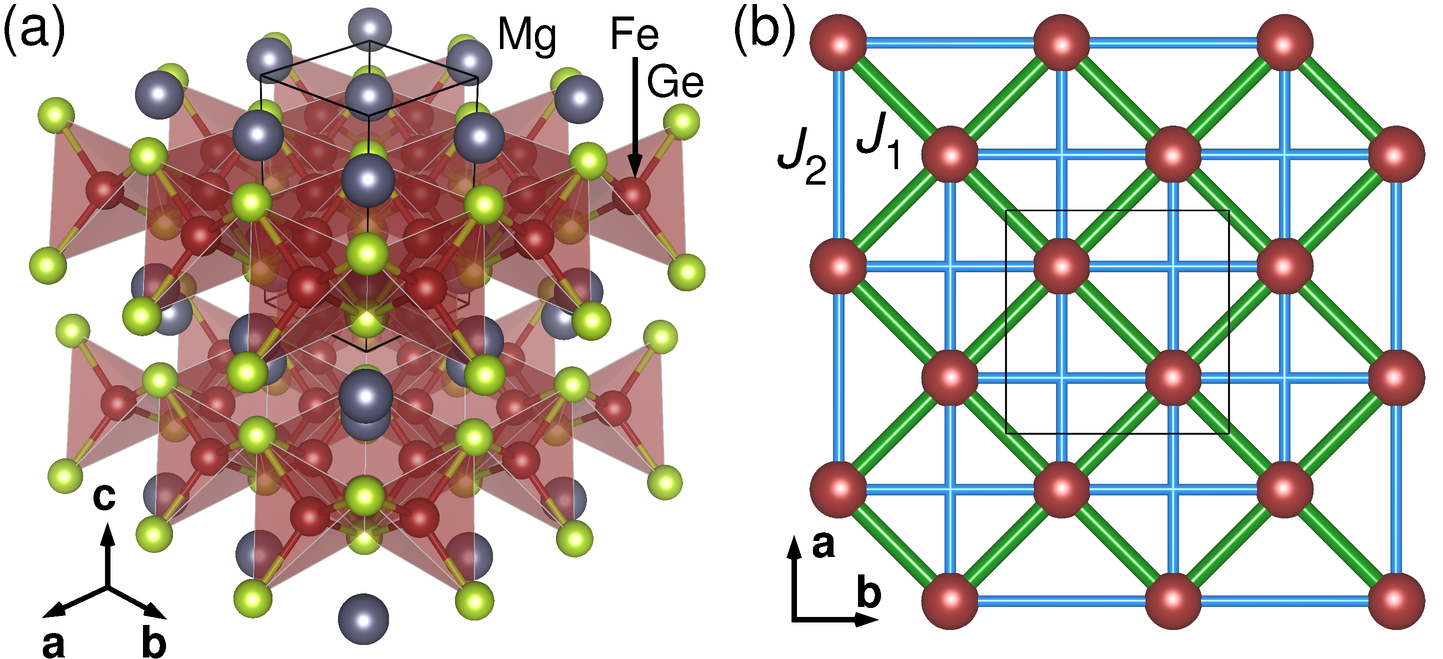}
\caption{(Color online) (a) Structure of MgFeGe.  (b) Exchange
  coupling paths $J_1$ and $J_2$ in the Fe plane.}
\label{fig:structure}
\end{figure}

The recently discovered MgFeGe compound~\cite{Liu2012}
(Fig.~\ref{fig:structure} (a)) is isostructural and isoelectronic to
the so-called 111 iron pnictides, LiFeAs and NaFeAs. In particular the
former is a good superconductor with $T_{c}=18$~K~\cite{Tapp2008},
and, like MgFeGe, is a paramagnetic metal in its normal
state. Moreover, the electronic structures of both compounds,
including their Fermi surfaces, calculated without any account of
magnetism (which seems logical, in view of the experimental
situation), are nearly identical~\cite{Liu2012,Rhee2013}, and so are
the calculated non-interacting susceptibilities. By implication, the
spin fluctuation spectra in both compounds must be also very close,
which seems, at first glance, to invalidate theories ascribing
superconductivity in iron pnictides to spin fluctuations.

In this paper we show that this is not the case. In fact, spin
fluctuations in the two materials are qualitatively different. The
result obtained by Rhee and Pickett~\cite{Rhee2013} is indeed at
variance with the popular weak coupling
scenario~\cite{Mazin2008,Hirschfeld2011,Chubukov2012}, which attempts
to describe these materials as nonmagnetic (not paramagnetic), in
terms of the linear response of the nonmagnetic Fermi surface,
however, in this case the weak coupling approach appears to be
inadequate.

The main problem with the density functional description of the
paramagnetic phases of Fe pnictides is that standard density
functional calculations cannot handle paramagnetism as disordered
local moments. The standard approach~\cite{Gyorffy1985} when
calculations are performed without allowing for a nonzero spin
density, simply forces each ion into a completely nonmagnetic
state. As discussed in numerous
papers~\cite{Yildirim2009,Mazin2009,Zhang2010, Opahle2009}, in all Fe
pnictides, with a notable exception of the collapsed tetragonal phase
in CaFe$_{2}$As$_{2},$ not only the local moments of the order of 2
$\mu_{B}$ remain, but they are apparently correlated in the standard
stripe manner. This is evidenced, for instance, by the fact that the
lattice dynamics and equilibrium structure of formally nonmagnetic
pnictides only agree with the experiment in spin-unrestricted
calculations with the full density functional theory (DFT) magnetic
moment of $\sim$ 2 $\mu_{B}$~\cite{Yildirim2009,Mazin2009,Zhang2010}.
Also, de Haas van Alphen experiments on BaFe$_{2}$As$_{2}$ are much
better described by DFT calculations with the full self-consistent
magnetization, rather than with the much smaller experimentally
reported magnetic moment~\cite{dHvA}. Computationally, the hallmark of
local moments is the possibility to converge calculations to different
magnetically ordered states, with the energy difference between them
considerably smaller than between any of them and the nonmagnetic
state.

With this in mind, we have deliberately stepped out of the weak
coupling domain and searched for magnetic solutions.  Note that DFT
{\it per se} is a mean field theory, but not a weak coupling theory.
The investigation of different magnetic configurations gives us a clue
of what kind of local correlations one can expect and what the
structure of the spin susceptibility, as opposed to weak coupling,
should be.

In most iron pnictides these two approaches (strong and weak coupling)
give the same result: spin fluctuations are peaked at ($\pi,\pi)$ in
the folded Brillouin zone. In iron selenides this is not exactly the
case. In the pure FeTe compound, the ground state (experimental and
calculated) is a so-called double stripe, corresponding, in the same
notation, to the $(\pi/2,0)$ ordering vector. However, the more
familiar stripe state is very close in energy, and when
superconductivity is suppressed by alloying with Se, not only is the
long-range order suppressed, but also, to the same degree, the local
moments. As a result, spin fluctuations corresponding to stripe-like
local correlations, reappear and gradually overcome the $(\pi/2,0)$
fluctuations. The former are pairing in the $s_{\pm}$ scenario, and,
in agreement with the concept, superconductivity appears.

Therefore, in order to properly compare LiFeAs and MgFeGe and to
understand the character of the spin fluctuations, we need to perform
a spin-unrestricted mean-field DFT calculation for both, and only if
the results will be reasonably close we can claim a failure of the
spin-fluctuation model.

\begin{table}
  \caption{Energies of various spin configurations
    with respect to the nonmagnetic solution for MgFeGe and LiFeAs, in meV/Fe}
\label{table}
\begin{ruledtabular}
\begin{tabular}{l|r|r|r|r}
&N{\'e}el &double stripe &stripe  & ferromagnetic \\\hline
 MgFeGe  &  -113&  -175  &  -179 &  -183 \\
 { LiFeAs} &   -56 &   -94 &   -133 &   -37
\end{tabular}
\end{ruledtabular}
\end{table}

\begin{figure}[ptb]
\includegraphics[width=.45\textwidth]{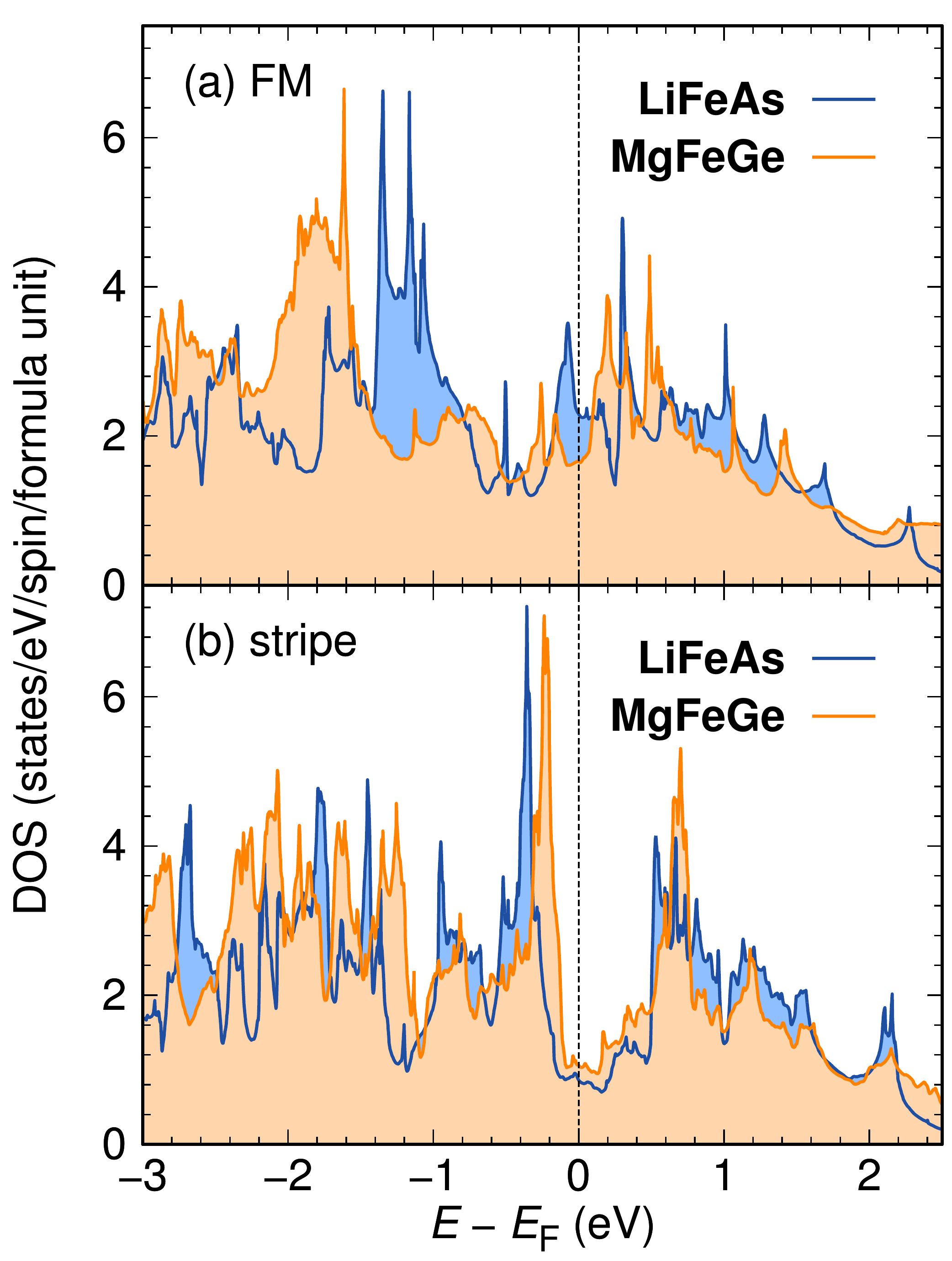}
\caption{(Color online) Comparison of densities of states for LiFeAs
  and MgFeGe in different magnetic configurations. (a), (b) LiFeAs in
  ferromagnetic and stripe-type antiferromagnetic order,
  respectively. (c), (d) MgFeGe in ferromagnetic and stripe-type
  antiferromagnetic order, respectively.}
\label{fig:dos}
\end{figure}

\begin{figure}[ptb]
\includegraphics[width=.3\textwidth]{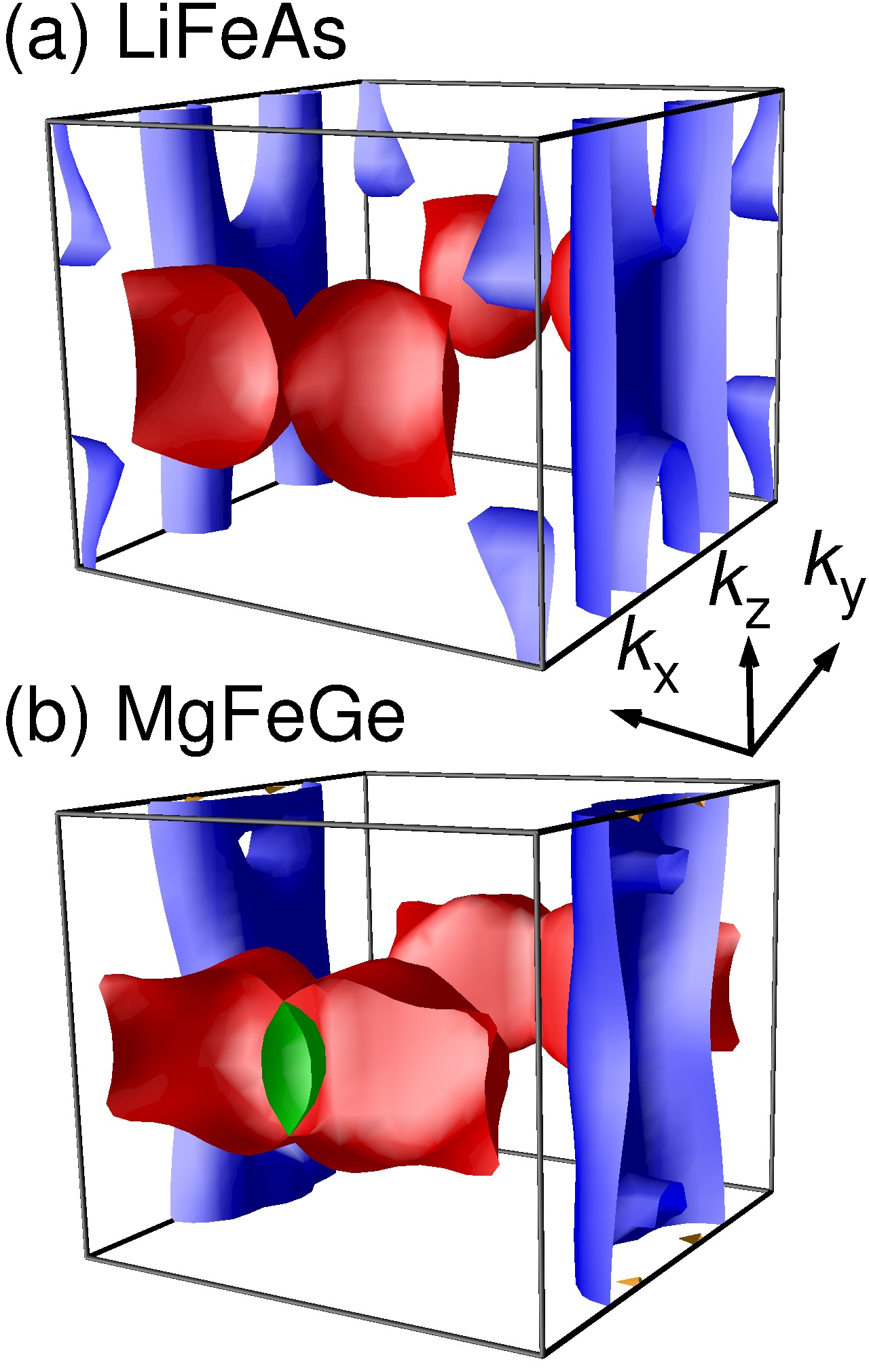}
\caption{(Color online) Fermi surfaces in the AF stripe configuration
  for (a) LiFeAs and (b) MgFeGe. The corners of the reciprocal lattice
  unit cell are at the $\Gamma$ points, $x$ and $y$ are the ferro- and
  antiferromagnetic directions, respectively.}
\label{fig:fs}
\end{figure}

\begin{figure}[ptb]
\includegraphics[width=.45\textwidth]{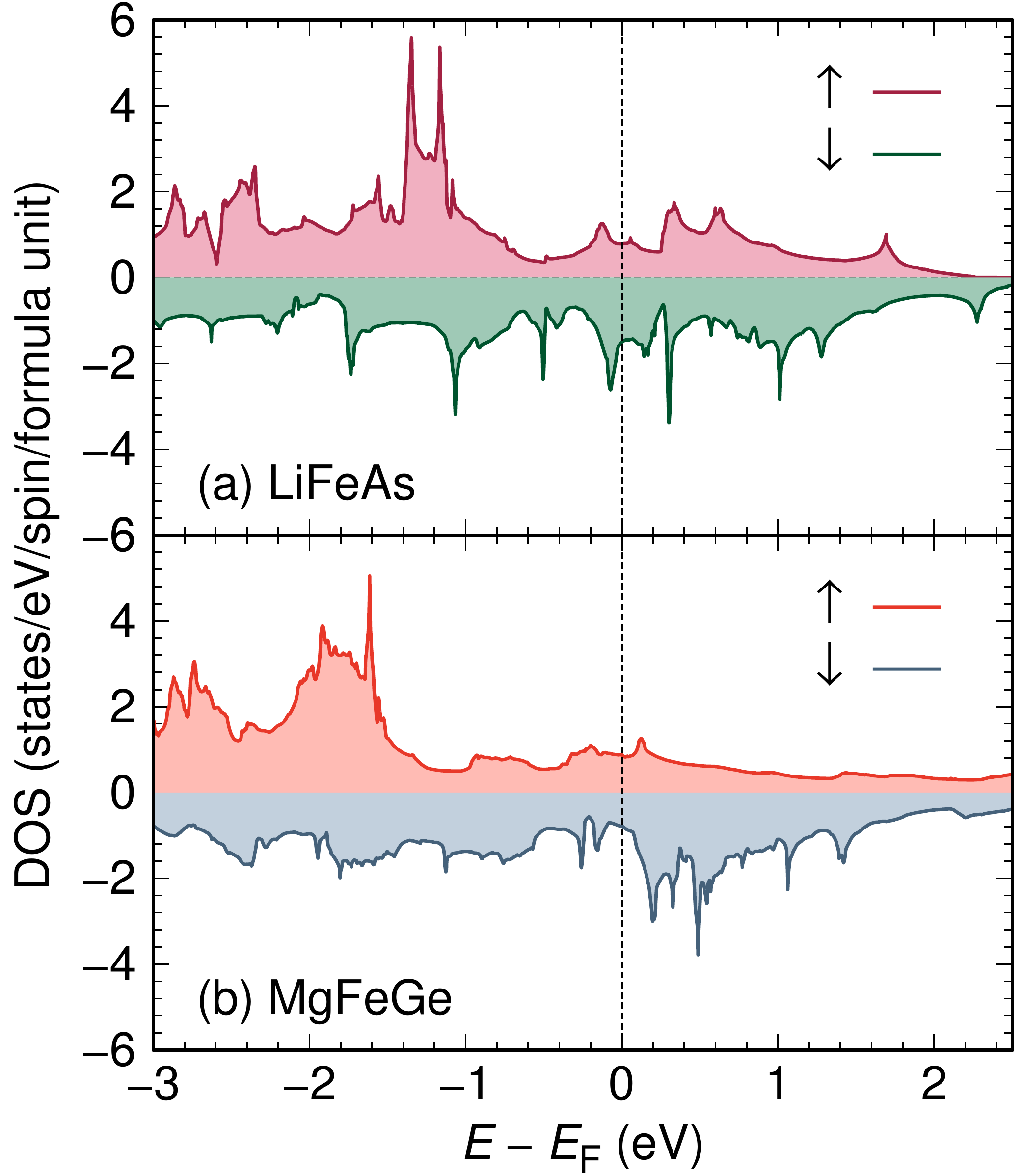}
\caption{(Color online) Densities of states in the FM configuration
  for (a) LiFeAs and (b) MgFeGe.}
\label{fig:fmdos}
\end{figure}

In Table~\ref{table}, we compare the energies of various
calculated~\cite{calc,w2k,FPLO} magnetic structures for LiFeAs and
MgFeGe\cite{noteH,Kresse}.  It has been pointed out previously that compared
to 1111 and 122 structures, in LiFeAs the ferromagnetic state is
relatively stable, even though still definitely
energetically above the antiferromagnetic stripe
phase.~\cite{Buechner,Qian} In MgFeGe this tendency is much stronger than
in LiFeAs and leads to an unexpected result: the calculated ground
state is actually ferromagnetic!  Moreover, one can quantify the
effect by extracting the nearest- and the next-nearest-neighbor
effective exchange constants.~\cite{comment_heis} For LiFeAs we find,
as expected, two antiferromagnetic exchange constants: $J_{1}=52$~meV
and $J_{2}=102$~meV corresponding to the Fe - Fe interaction paths
shown in Fig.~\ref{fig:structure}.  The fact that $J_{2}>J_{1}/2$
reflects the stripe phase being the ground state. For MgFeGe we obtain
$J_{1}=-71$~meV and $J_{2}=32$~meV. Note that now the nearest neighbor
interaction is ferromagnetic. Incidentally, this fact invalidates the
set of theories explaining magnetic properties of Fe-based
superconductors in terms of conventional superexchange; indeed,
superexchange in this geometry can only be antiferromagnetic,
therefore the nature of the ferromagnetic interaction in MgFeGe must
be of more complex, and, as we will see next, of itinerant
origin.~\cite{comment_moment}

The origin of this phenomenon can be understood by comparing the
density of states (DOS) of the two compounds in ferromagnetic and
stripe-type antiferromagnetic configurations (see Fig.~\ref{fig:dos}).
We observe that in LiFeAs the DOS in the stripe phase is drastically
reduced with respect to the DOS in the ferromagnetic phase and a
pseudogap forms around the Fermi energy with a substantial gain of the
one-electron energy. In MgFeGe, on the other hand, ferromagnetic and
stripe order lead to similar DOS at the Fermi energy and
ferromagnetism wins by a small energy amount (see Table~\ref{table}).
Also an analysis of the Fermi surfaces of the two compounds in the
antiferromagnetic stripe configuration support these results (see
Fig.~\ref{fig:fs}).  Indeed, while in the nonmagnetic phase, as shown
in Ref.~\onlinecite{Rhee2013}, the Fermi surfaces of the two compounds
are nearly identical, in the stripe phase of LiFeAs, just as in all
other Fe pnictides, the Fermi surface is mostly gapped.  In MgFeGe, on
the other hand, the gapping is much smaller, and so is the
corresponding energy gain.

One can also ask another question: is this difference due to different
crystallographic parameters or different ionic properties? To answer
this question, we have performed calculations for a hypothetical
compound with the composition of MgFeGe, but crystallographic
parameters as in LiFeAs.  Interestingly, we found that the energy
difference between the stripe and the ferromagnetic states was
strongly reduced (by about a factor of 7), but the sign was still the
same, favoring the ferromagnetic order. This indicates that both
chemistry $and$ crystallography contribute to the difference in
magnetic properties between the two compounds.

To summarize, by investigating the magnetic behavior of MgFeGe versus
LiFeAs we found that MgFeGe is most stable in a ferromagnetic
configuration in contrast to LiFeAs that stabilizes in the more
familiar antiferromagnetic stripe-like pattern. This has important
consequences for the actual behavior of MgFeGe: the short range
correlations, and, by implication, fluctuations in the paramagnetic
state, are stronger at $q=0$ than at $\mathbf{q=}(\pi,\pi).$ Thus,
they actually destroy, rather than support the $s_{\pm}$ pairing. The
fact that MgFeGe is not superconducting therefore supports the
spin-fluctuation induced pairing model and the $s_{\pm}$ pairing
state. Another, probably more critical message is that although many
useful results have been obtained using the weak coupling linear
response methodology, this path is slippery.  Neglecting the fact that
DFT calculations, as well as certain experiments, for iron based
superconductors point toward strong coupling and large local moments
may be dangerous.

The last note concerns the role of correlations in MgFeGe. Inclusion
of correlation effects beyond DFT in
LiFeAs~\cite{Yin2011,Ferber2012a,Ferber2012b} yields a better
description of its Fermi surface in agreement with experimental
observations.  MgFeGe is probably more strongly correlated than
LiFeAs, since it has a larger magnetic moment (compare in Fig.~
\ref{fig:fmdos} the up and down DOS).  By the same argument, it is
probably less correlated than FeSe, and thus correlation effects are
unlikely to be responsible for the absence of superconductivity, as
opposed to proximity to the ferromagnetic instability, which does not
appear in any superconducting Fe pnictide or chalcogenide.

I.I.M. acknowledges support from the Funding from the Office of Naval
Research (ONR) through the Naval Research Laboratory's Basic Research
Program, and of the Alexander von Humboldt Foundation. H.O.J. and
R.V. acknowledge support by the Deutsche Forschungsgemeinschaft (DFG)
through SPP 1458. We thank W. Pickett for pointing out to us the
existing controversy on MgFeGe.


\begin{thebibliography}{9}
\bibitem{Gyorffy1985} Note that we are not considering here such
  modifications as the disordered-local-moments model proposed by
  B. L. Gyorffy, A. J. Pindor, J. Staunton, G. M. Stocks and
  H. Winter, J. Phys. F: Metal Phys. {\bf 15}, 1337 (1985).
\bibitem{comment_heis} Note that the estimation of the nearest and
  next nearest neighbor exchange constants doesn't mean that we
  endorse using the short-range Heisenberg model as a physically
  meaningful description of these itinerant systems; it is well known
  that proper mapping of the DFT energy differences requires not only
  going beyond the second coordination shell, but even beyond the
  Heisenberg model.
\bibitem{comment_moment} Note that a trivial, but often overlooked
  fact is that local magnetic moments do not necessarily imply
  localized electrons; for instance, metallic iron has itinerant
  electrons and local moments.
\end{thebibliography}
\end{document}